\documentstyle[12pt]{article}
\def\beq{\begin{equation}}
\def\eeq{\end{equation}}

\def\al{\alpha}

\def\ga{\gamma}

\def\de{\delta}

\def\si{\sigma}
\def\Si{\Sigma}

\def\tm{\times}

\def\lam{\lambda}

\def\om{\omega}
\def\ep{\epsilon}

\def\sq{\sqrt}

\def\l{\left (}
\def\r{\right )}
\def\fr{\frac}
\def\la{\label}
\def\hs{\hspace}

\def\ran{\rangle}
\def\lan{\langle}
\def\ov{\overline}

\begin{document}
\begin{titlepage}

\begin{center}   
{\Large\bf 
5D  SUSY Orbifold SU(6) GUT and \\
Pseudo-Goldstone Higgs Doublets}  
\end{center} 
\vspace{0.5cm} 
\begin{center} 
{\large Filipe \hs{-0.07cm}Paccetti \hs{-0.07cm}Correia$^{a}$\footnote{E-mail address: 
F.Paccetti@ThPhys.Uni-Heidelberg.DE} , 
{}Michael \hs{-0.07cm}G. \hs{-0.07cm}Schmidt$^{a}$\footnote{E-mail address: 
M.G.Schmidt@ThPhys.Uni-Heidelberg.DE} , 
{}Zurab \hs{-0.07cm}Tavartkiladze$^{a, b}$\footnote{E-mail address: 
Z.Tavartkiladze@ThPhys.Uni-Heidelberg.DE} 
} 
\vspace{0.5cm}

$^a${\em Institut f\"ur Theoretische Physik, Universit\"at Heidelberg,
Philosophenweg 16,\\
 D-69120 Heidelberg, Germany \\

$^b$ Institute of Physics, 
Georgian Academy of Sciences, Tbilisi 380077, Georgia\\

} 

\end{center}

\vspace{1.0cm}

\begin{abstract}

We construct a 5D SUSY $SU(6)$ GUT on an $S^{(1)}/Z_2\tm Z_2'$ orbifold.
The first stage of gauge symmetry breaking occurs through compactification
and a
specific selection of boundary conditions. Additional symmetries play
a crucial role for the generation of $\mu $ and $B\mu $ terms of
appropriate values:
with a $SU(2)_{cus}$ custodial  symmetry the Higgs
doublets
naturally emerge as massless pseudo-Goldstone bosons in the unbroken SUSY
limit. After SUSY breaking they get masses of the order of the weak
scale. If
instead of $SU(2)_{cus}$ a discrete $Z_5$ symmetry is applied the Higgs
doublet's masses are still adequately suppressed, but they are not
pseudo-Goldstones. The $Z_5$ discrete symmetry also can be very
important for
GUT scale generation and an all order hierarchy. Fermion masses are
naturally
generated and nicely blend with additional symmetries. In
the considered scenario unification of the
three gauge couplings occurs near $10^{16}$~GeV.

\end{abstract}

\end{titlepage}

\section{Introduction}
The doublet-triplet (DT) splitting problem is an 'Achilles heel' of any
GUT. When constructing a realistic model, one has to
explain why the doublet components are split in mass from their
colored triplet partners, with a ratio 
$M_T/M_D\stackrel{>}{_\sim} 10^{13}$. A big triplet mass is needed for
the
proton stability and for maintaining successful unification of the three
gauge coupling constants. Recently a way was proposed, based on 
a higher dimensional construction \cite{kawa},
\cite{dtpdecay}-\cite{orbSO10} how to cure various 
phenomenological problems common to GUTs in an economical way: 
It was observed that the compactification of a 
5D ($N=1$) SUSY $SU(5)$ model on an $S^{(1)}/Z_2\tm Z_2'$ orbifold 
provides a natural DT splitting,
GUT symmetry breaking and proton stability. Furthermore, due to 5D 
$N=1$ SUSY, the $\mu $-term vanishes at the 5D level. This can be 
considered as an excellent starting point for realistic model building. 
However, on the 4D level, there
is no reason to have a vanishing $\mu $-term and some specific extensions
\cite{mu1}-\cite{mu3} are required to provide a $\mu $-term of
appropriate
magnitude. One of
the attractive ideas for the solution of DT splitting and $\mu $ 
problems in 4D is the
pseudo-Goldstone boson (PGB) mechanism \cite{pgbsu61}-\cite{radbr_pgb},
where Higgs doublets
emerge as Goldstone modes of a spontaneously broken global symmetry. If
the symmetry of the 'scalar' superpotential is larger than the symmetry of
the whole Lagrangian, then due to the symmetry breaking PGBs emerge, which
are
massless, as stated in the Goldstone theorem. Thanks to SUSY, the
superpotential is not
renormalized and the PGBs remain massless in the unbroken SUSY limit. 
After the SUSY
breaking their masses will not exceed the SUSY scale $m\sim 1$~TeV. In
this way the DT splitting problem can be resolved. It also turns out that
the $\mu $- and 
$B\mu $-terms get the right magnitude.

In this paper we consider a SUSY $SU(6)$ GUT in five dimensions. The first
stage of GUT symmetry breaking occurs via compactification of one extra
dimension, leaving an unbroken $SU(3)_c\tm SU(3)_L\tm U(1)$ symmetry on the 4D
level. Then we invoke the custodial $SU(2)_{cus}$ symmetry \cite{cus}
in order to get a natural solution of the $\mu $ problem. 
The breaking of $SU(3)_L$
to $SU(2)_L$ by the Higgs mechanism requires the existence of 
Goldstone doublets in the Higgs sector. 
Although these are eaten by the massive gauge bosons, 
there are additional states related to them by the $SU(2)_{cus}$ symmetry.
These will remain massless PGBs in the unbroken SUSY regime and 
are good candidates for the MSSM doublet-antidoublet
pair. After SUSY
breaking the $B\mu $ and $\mu $ terms are naturally generated and have
values of the order of the SUSY scale. 
We emphasize that some difficulties, present in 
4D $SU(6)$ GUTs with custodial $SU(2)_{cus}$ symmetry mechanism \cite{cus} 
such as the
emergence of an undesirable intermediate scale (which disrupts the
unification of the three gauge coupling constants) do not exist in our model due to
the extra-dimensional construction. The model is very
economical as there is no need for further extensions \cite{cusext} in
order
to avoid the difficulties mentioned before and to make the $SU(2)_{cus}$
symmetry mechanism \cite{cus} efficient.
We also present a discussion about the emergence of the $SU(2)_{cus}$
symmetry in the theory. An alternative model is constructed, in which
$SU(2)_{cus}$ is replaced by a ${\cal Z}_5$ discrete symmetry. The latter
provides an all order hierarchy and natural generation of the GUT scale
$M_G\sim 10^{16}$~GeV. The generation of fermion masses and the 
unification of the
gauge coupling constants are also studied.
\section{5D SUSY $SU(6)$ on $S^{(1)}/Z_2\tm Z_2'$ orbifold}
Consider a 5D $N=1$ SUSY $SU(6)$ GUT where the 5th dimension is a $S^{(1)}/Z_2\tm Z_2'$ orbifold. By a suitable assignment of the orbifold $Z_2\tm Z_2'$ parities, $(P, P')=(\pm , \pm)$, to the different components
of the gauge fields, it is possible to break the gauge
symmetry and half of the supersymmetries at a fixed point. 
The 5D $N=1$ gauge
supermultiplet is, from a $4D$ point of view, a $N=2$ supermultiplet
$V_{N=2}$, which contains a 4D $N=1$ gauge superfield $V$ and a chiral superfield
$\Si$, both in the adjoint representation ${\bf 35}$ of $SU(6)$:
$V_{N=2}=(V, ~\Si )$. We select boundary conditions in such a way as to
break $SU(6)$ to $SU(3)_c\tm SU(3)_L\tm U(1)\equiv G_{331}$. 
It will turn out that this breaking
(through the $G_{331}$ channel) is crucial if one wants to obtain the 
MSSM Higgses as PGB modes
after a second stage of symmetry breaking. In terms of $G_{331}$ the
decomposition of $V(35)$ reads
\beq
V(35)=V_c(8, 1)_0+V_{3L}(1, 8)_0+V_{U(1)}(1, 1)_0+
V_{c\bar L}(3, \ov{3})_2+V_{\bar c L}(\ov{3}, 3)_{-2}~,
\la{dec35}
\eeq
and similar for $\Si (35)$. The subscripts in eq.(\ref{dec35}) denote
$U(1)$ hypercharges in $1/\sq{12}$ units:
\beq
Y_{U(1)}=\fr{1}{\sq{12}}\l 1,~1,~1,~-1,~-1,~-1 \r~.
\la{u1hyp}
\eeq
We choose the $Z_2\tm Z_2'$ parities of the fragments of $V$ and $\Si $ 
as
$$
\l V_c, V_{3L}, V_{U(1)}\r \sim (+, +)~,~~~
\l V_{c\bar L}, V_{\bar c L}\r\sim (-, +)~,
$$
\beq
\l \Si_c, \Si_{3L}, \Si_{U(1)}\r \sim (-, -)~,~~~
\l \Si_{c\bar L}, \Si_{\bar c L}\r\sim (+, -)~,
\la{gaugepar}
\eeq
so that at the $y=0$ fixed point we have 4D $N=1$ SUSY and $G_{331}$ gauge
symmetry.

\subsection{$G_{331}$ breaking and pseudo-Goldstone Higgses}

{}For the further breaking of $G_{331}$ down to the standard model group
$SU(3)_c\tm SU(2)_L\tm U(1)_Y\equiv G_{321}$, we introduce the
following Higgses on the 5D level 
\beq
{\cal H}_{N=2}^{(i)}(6)=({\cal H},~\ov{\cal H})^{(i)}~,~~~
{{\cal H}_{N=2}'}^{(i)}(6)=({\cal H}',~\ov{\cal H}\hs{0.5mm}')^{(i)}~,~~~
i=1,2~,
\la{higgses}
\eeq
where ${\cal H}^{(i)}$, ${{\cal H}'}^{(i)}$ and $\ov{\cal H}^{(i)}$,
${\ov{\cal H}\hs{0.5mm}'}^{(i)}$ are in the representations ${\bf 6}$ and 
${\bf \ov{6}}$ of $SU(6)$, resp. In terms of $G_{331}$, 
${\cal H}^{(i)}$ and ${\ov{\cal H}}^{(i)}$ decompose as 
\beq
{\cal H}^{(i)}=T^{(i)}(3, 1)_1+H^{(i)}(1, 3)_{-1}~,~~~
\ov{\cal H}^{(i)}=\ov{T}^{(i)}(\ov{3}, 1)_{-1}+
\ov{H}^{(i)}(1, \ov{3})_{1}~,
\la{dechiggs}
\eeq
and similar for ${{\cal H}'}^{(i)}$ and ${\ov{\cal H}\hs{0.5mm}'}^{(i)}$. 
With the $Z_2\tm Z_2'$ parities
$$
(H^{(i)},~{\ov{H}\hs{0.5mm}'}^{(i)})\sim (+,~+)~,~~~
(\ov{H}^{(i)},~{H\hs{0.5mm}'}^{(i)})\sim (-,~-)~,
$$
\beq
(T^{(i)},~{\ov{T}\hs{0.5mm}'}^{(i)})\sim (-,~+)~,~~~
(\ov{T}^{(i)},~{T\hs{0.5mm}'}^{(i)})\sim (+,~-)~,
\la{parhiggs}
\eeq
we have two $SU(3)_L$ triplet-antitriplet pairs 
$H^{(i)}$, ${\ov{H}\hs{0.5mm}'}^{(i)}$ at the $y=0$ fixed point . 
The $SU(3)_c$ triplets 
$T^{(i)}$, ${T\hs{0.5mm}'}^{(i)}$, $\ov{T}^{(i)}$, ${T\hs{0.5mm}'}^{(i)}$
have no zero modes. The $H$ fields can be used to break $SU(3)_L\tm U(1)$ down
to $SU(2)_L\tm U(1)_Y$. Note that for this purpose it is 
sufficient to have one $SU(3)_L$ triplet-antitriplet pair, which we choose
to be
$H^{(1)}$, ${\ov{H}\hs{0.5mm}'}^{(1)}$. If the third components of
these triplets get non zero 'scalar' VEVs, then the breaking
$SU(3)_L\tm U(1)\to SU(2)_L\tm U(1)_Y$ occurs. Consequently,
the $SU(2)_L$ doublet and antidoublet components of $H^{(1)}$ and
${\ov{H}\hs{0.5mm}'}^{(1)}$ resp. are absorbed by the appropriate gauge
fields and therefore are genuine Goldstones. But we have an additional
pair of states, $H^{(2)}$, ${\ov{H}\hs{0.5mm}'}^{(2)}$, with precisely the same
transformation properties as $H^{(1)}$, ${\ov{H}\hs{0.5mm}'}^{(1)}$. If 
the $H^{(2)}$ and ${\ov{H}\hs{0.5mm}'}^{(2)}$ states are related to the
 $H^{(1)}$ and ${\ov{H}\hs{0.5mm}'}^{(1)}$ resp. by some specific
symmetry, then the unabsorbed physical doublets might be  also massless. 
If some care is taken in
this procedure and they are massless at the tree level, then due to SUSY
they will be
protected from getting a radiative mass and turn out to be massless
pseudo-Goldstones in the unbroken SUSY limit.
To realize this, we introduce a custodial symmetry, $SU(2)_{cus}$, under which
$H^{(1)}$, $H^{(2)}$ form a doublet $(H^{(1)},~H^{(2)})\equiv H_m$
[where $m=1, 2$ is $SU(2)_{cus}$ index] and 
${\ov{H}\hs{0.5mm}'}^{(1)}$, ${\ov{H}\hs{0.5mm}'}^{(2)}$ - an antidoublet
$({\ov{H}\hs{0.5mm}'}^{(1)},~{\ov{H}\hs{0.5mm}'}^{(2)})\equiv \ov{H}^m$.
The component structure of $\ov{H}^m$, $H_m$ can schematically be 
represented as
\beq
\begin{array}{cc}
\vspace{2mm}
H_m=
\begin{array}{c}
 \end{array}\!\!\! &{\left(\begin{array}{cc}
D_1  &D_2 
\\
\chi_1   &\chi_2    
\end{array}\right) }~,
\end{array}  \!\!  ~~~~~~~
\begin{array}{cc}
  \vspace{2mm}
\ov{H}^m =
{\left(\begin{array}{cc}
\ov{D}^1   &\ov{D}^2
\\
\ov{\chi }^1  &\ov{\chi }^2
\end{array}\right) }~.
\end{array}  \!\!  
\label{structH}
\end{equation}
where $D$, $\ov{D}$ are doublet-antidoublets of $SU(2)_L$ and
$\chi $, $\ov{\chi }$ (third components of $H$, $\ov{H}$) are $SU(2)_L$
singlets. With VEVs
\beq
\begin{array}{cc}
\vspace{2mm}
\lan H_m\ran =
\begin{array}{c}
 \end{array}\!\!\! &{\left(\begin{array}{cc}
0 & 0
\\
v_0   &0    
\end{array}\right) }~,
\end{array}  \!\!  ~~~~~~~
\begin{array}{cc}
  \vspace{2mm}
\lan \ov{H}^m\ran  =
{\left(\begin{array}{cc}
0  &0
\\
v_0  &0
\end{array}\right) }~,
\end{array}  \!\!  
\label{vevH}
\end{equation}
the $SU(3)_L\tm U(1)$ will be broken down to $SU(2)_L\tm U(1)_Y$, where
\beq
Y=\fr{2}{\sq{5}}Y_{U(1)}-\fr{1}{\sq{5}}Y_{U(1)'}~.
\la{SMhyp}
\eeq
$Y_{U(1)}$ is given in (\ref{u1hyp}) and $Y_{U(1)'}$ corresponds to the 
$SU(3)_L$'s broken generator
\beq
Y_{U(1)'}=\fr{1}{\sq{12}}{\rm Diag}\l 1,~1,~-2\r~.
\la{brY}
\eeq

Let us now demonstrate more precisely how the mechanism works. For this purpose we will
write 4D brane couplings. We also introduce a singlet superfield $S$ with
zero mode state\footnote{On the 5D level one can have a singlet state 
${\cal S}_{N=2}=(S,~\ov{S})$ and the following orbifold parities for its 
fragments: $S\sim (+, +)$, $\ov{S}\sim (-, -)$.}.
As the relevant renormalizable 4D 'scalar' superpotential, invariant under
the $G_{331}\tm SU(2)_{cus}$ symmetry, we take
\beq
W=\ov{H}^m\l M_H+\lam S\r H_m+\fr{1}{2}M_SS^2+\fr{1}{3}\si S^3~,
\la{4Dsup}
\eeq
where $M_H$, $M_S$ are mass scales and $\lam$, $\si $ are dimensionless
couplings.
With (\ref{vevH}) all $D$-terms automatically vanish, while the conditions 
$F_{\ov{H}^m}=F_{H_m}=F_S=0$ give us
\beq
\lan S \ran\equiv S_0=-\fr{M_H}{\lam }~,~~~~
v_0^2=\fr{M_H}{\lam}\l M_S-\fr{\si }{\lam }M_H\r~.
\la{solVEV}
\eeq
With (\ref{solVEV}) and (\ref{structH}), (\ref{vevH}) it is easy to see
from (\ref{4Dsup}) that the superpotential masses of $D_m$, $\ov{D}^m$ doublet
superfields vanish: $M_D=M_H+\lam \lan S\ran =0$. Let's consider the mass
spectrum for the scalar and the fermionic doublet components separately. 
As we have
already seen, none of them gets mass from the superpotential. On the other hand
we must take into account the $SU(3)_L$ $D$-terms contribution to the Higgs
doublets potential. The state $\fr{1}{\sq{2}}(D_1+\ov{D}^{1*})_s$ [the subscript 's'
refers to the scalar component of appropriate superfield] is the 
massless genuine Goldstone boson absorbed by the doublet-antidoublet gauge bosons, while the states $\fr{1}{\sq{2}}(\ov{D}^{1*}-D_1)_s$ 
acquire the mass
$g_{SU(3)_L}v_0$ from the $D$-terms ($g_{SU(3)_L}$ denotes the $SU(3)_L$
gauge
coupling constant). Of course the same happens with their fermionic
partners: they
get the mass $g_{SU(3)_L}v_0$ through the mixing with the $SU(2)_L$
doublet-antidoublet gauginos. This is indeed the supersymmetric Higgs
mechanism (i.e. the discussion can be applied to complete superfields,
also
including their auxiliary fields). As far as the second pair of
doublets, $D_2$, $\ov{D}^2$, is
concerned, they are related to $D_1$, $\ov{D}^1$ by the $SU(2)_{cus}$ symmetry
and thus their masses from the 
superpotential (\ref{4Dsup}) are zero. Therefore, we will identify them
with the MSSM pair of doublets
\beq
D_2\equiv h_u~,~~~~~\ov{D}^2\equiv h_d~.
\la{MSSMpair}
\eeq
Note that $h_u$ and $h_d$ are massless PGBs. For this, the $SU(2)_{cus}$
symmetry
is crucial,
which forbids the mixing terms 
$M_{12}\ov{H}^1H_2$, $M_{21}\ov{H}^2H_1$ in the superpotential. These
couplings would destroy the hierarchy since  
$M_{12}\sim M_{21}\gg {\cal O}(100~{\rm GeV})$.

Before we turn our attention to the issue of SUSY breaking, let us comment on
the singlet states coming from $H_m$,
$\ov{H}^m$ [see (\ref{structH})] and $S$.
The scalar
component ${\rm Im}(\ov{\chi}^1-\chi_1)_s$ is a genuine Goldstone
eaten up by the gauge field which corresponds to the broken Abelian
factor, while the superposition 
${\rm Re}(\chi_1-\ov{\chi }^1)_s$ gets mass 
$2g_{SU(3)_L}^2v_0^2$ from the $D$-terms. 
Other scalar singlets from $\ov{\chi }^1$, $\chi_1$ and $S$ get masses
of the order of $M_G$.
The fermionic
superpartner's 
mass spectrum looks similarly due to SUSY. 
As far as the states $\chi_2$, $\ov{\chi }^2$ are concerned,
they remain massless in unbroken SUSY limit.
This spectrum of particles can be also explained with symmetry arguments.
Looking on (\ref{4Dsup}) one can verify that the $SU(2)_{cus}$ symmetry,
which
acts on $\ov{H}^m$, $H_m$, gives rise to an accidental $U(6)_{Gl}$
symmetry
in
the superpotential. The VEVs in (\ref{vevH}) break $U(6)_{Gl}$ down to
$U(5)_{Gl}$ and the number of Goldstones is $N_{Gol}=36-25=11$. At the
same
time, the same VEVs break the $SU(3)_L\tm U(1)$ local symmetry down to
$SU(2)_L\tm U(1)_Y$ and therefore the number of genuine Goldstones is
$N_{GGol}=8+1-(3+1)=5$. So, from eleven($=N_{Gol}$) Goldstones only
five($=N_{GGol}$) are absorbed and the remaining six are
pseudo-Goldstones.
Amongst them four correspond to the doublet-antidoublet pair $h_u$,
$h_d$ and the remaining two are the MSSM singlets $\chi_2$, $\ov{\chi
}^2$. Of
course, the $D$-terms do not have $U(6)_{Gl}$ symmetry, but they do not
contribute to the pseudo-Goldstone masses unless SUSY is broken.

{\bf SUSY breaking: generation of $B\mu $ and $\mu $}

Throughout our discussions above we assumed that SUSY is
unbroken. We now investigate the effects of SUSY breaking.
As we will see, the inclusion of soft SUSY breaking terms
automatically generates $B\mu $ and $\mu $ terms of the order of the SUSY
breaking scale ($\sim m^2$ and $m$ resp.). 
Let us thus take all soft
SUSY breaking terms (for $H_m$, $\ov{H}^m$ and $S$) in the potential
$$
V_{soft}=m_{\ov{H}}^2|\ov{H}^m|^2+m_{H}^2|H_m|^2+m_s^2|S|^2+
A_1m[\ov{H}^m(M_H'+\lam S)H_m+{\rm h.c.}]+
$$
\beq
A_2m\l \fr{1}{2}M_S'S^2+\fr{1}{3}\si S^3+{\rm h.c.}\r ~,
\la{soft}
\eeq
[In (\ref{soft}) all fields denote the scalar components of the associated
superfields].
(\ref{soft}) is the most general soft SUSY 
breaking potential,
where neither universality nor proportionality are assumed. The masses
$m_{\ov{H}}$, $m_{H}$, $m_s$ are of the order of the SUSY scale $m\sim 1$~TeV,
while $M_H'\sim M_H$ and $M_S'\sim M_S$.
We also assume that the dimensionless couplings $A_1$, $A_2$ are of the
order
of one.
The complete potential has the form
\beq
V=|F_{\ov{H}^m}|^2+|F_{H_m}|^2+|F_S|^2+V_{soft}+D-{\rm terms}~,
\la{pot}
\eeq
where the $F$-terms must be derived from superpotential (\ref{4Dsup}).
The inclusion of soft terms will only slightly shift the solutions
(\ref{solVEV}), because we assume that $M_H, M_S\gg m$. 
We will search, therefore, for VEV solutions of
extrema of (\ref{pot}), in the form
\beq
\lan S\ran =S_0(1+x)~,~~~
\lan \chi_1 \ran \equiv v=v_0(1+y)~,~~~
\lan \ov{\chi }^1\ran \equiv \bar v=v_0(1+z)~,
\la{vev1}
\eeq
where $S_0$, $v_0$ are given in (\ref{solVEV}) and $x, y, z\ll 1$.
Different soft
masses $m_{H}^2$, $m_{\ov{H}}^2$ cause different values for
$v$, $\bar v$. Due to this, $D$-terms will not vanish anymore.   
In fact, $x, y, z$ must be expressed as powers
of $m/M$, where $M$ is a mass scale close to $M_H$,
$M_S$. After a straightforward analysis we find
$$
x=\fr{m}{2M_H}\l A_1-\fr{A_1}{\lam v_0^2}(M_S+2\si S_0)(M_H'+\lam S_0)+
\fr{A_2}{\lam v_0^2}(M_S'S_0+\si S_0^2)\r+\cdots ~,
$$ 
$$
y=-\fr{S_0}{2\lam v_0^2}(M_S+2\si S_0)x-
\fr{A_1}{2\lam^2v_0^2}(M_H'+\lam S_0)m+
\fr{m_{\ov{H}}^2-m_H^2}{8\ov{g}^2v_0^2}
+\cdots ~,
$$
\beq
z=-\fr{S_0}{2\lam v_0^2}(M_S+2\si S_0)x-
\fr{A_1}{2\lam^2v_0^2}(M_H'+\lam S_0)m-
\fr{m_{\ov{H}}^2-m_H^2}{8\ov{g}^2v_0^2}
+\cdots ~,
\la{solxy}
\eeq
where $\ov{g}^2=\fr{g_{3L}^2}{3}+\fr{g_1^2}{4}$ and dots in
(\ref{solxy}) stand for higher powers of $m/M$ which are
irrelevant for our studies. Since after SUSY breaking the VEV of $S$ is
shifted by the amount $S_0x$, in the superpotential (\ref{4Dsup}) there is no
precise
cancellation of the physical Higgs doublet superfield [defined by
(\ref{MSSMpair})] masses anymore. This
causes the generation of a $\mu $ term 
\beq
\mu=\lam S_0x=-M_Hx\sim m~.
\la{mu}
\eeq
{}For the same reason a $B\mu $ term is generated. From (\ref{pot}),
taking
into account (\ref{solVEV}), ({\ref{vev1}}), ({\ref{solxy}}) we obtain
\beq
B\mu =-\fr{1}{2}(m_H^2+m_{\ov{H}}^2)-M_H^2x^2 \sim m^2~.
\la{Bmu}
\eeq
Also, the states ${h_u}_s$, ${h_d}_s$ get 'direct' masses through
(\ref{pot}). 
The mass matrix for scalar components of $h_u$, $h_d$ is
\beq 
\begin{array}{cc}
 & {\begin{array}{cc}
\hs{-2.8cm}{h_u}_s^*&{h_d}_s
\end{array}}\\  
\vspace{2mm}
\begin{array}{c}
~~~~~~{h_u}_s\\
~~~~~~{h_d}_s^*
\end{array}
{\left(\begin{array}{cc}
m_1^2 &m_3^2 
\\
\hs{-0.1cm}m_3^2  &~m_2^2    
\end{array}\right) }~,
\end{array}  \!\!  ~~~~~~~
\label{higsmat}
\end{equation} 
where
$$
m_1^2=m_H^2+M_H^2x^2-2\ov{g}^2v_0^2(z-y)~,
$$
\beq
m_2^2=m_{\ov{H}}^2+M_H^2x^2+2\ov{g}^2v_0^2(z-y)~,~~~
m_3^2=B\mu~.
\la{higgsmass}
\eeq
are of the order of $m^2$.

If SUSY breaking occurs through minimal $N=1$ SUGRA \cite{sugra}  at
tree
level we have $m_H^2=m_{\ov{H}}^2$ and according to
(\ref{solxy}) $y=z$. This
means that $D$-terms do not contribute in Higgs masses and from
(\ref{Bmu}), (\ref{higgsmass}) we have $m_1^2=m_2^2=-m_3^2$ which means
that the determinant of matrix (\ref{higsmat}) is zero.
The state
$\fr{1}{\sq{2}}(h_d^*-h_u)_s$
has mass $2m_1^2\sim m^2$, while the superposition 
$\fr{1}{\sq{2}}(h_u+h_d^*)_s$ is massless at the tree level. The latter will
get a mass through radiative corrections \cite{radbr}, with the dominant
contributions coming from top-stop loops. 
Radiative electro-weak  symmetry breaking (EWSB) within PGB scenarios
where studied
in \cite{radbr_pgb}.
In our scenario due to a peculiarity of the model,
a shift between soft masses $m_H^2$ and $m_{\ov{H}}^2$ is already
sufficient to have an appropriate
EWSB. The reason for this is that the difference of $m_H^2$ and
$m_{\ov{H}}^2$ causes $y\neq z$ and $D$-terms contribute to the Higgs
masses. However, taking into account the solutions (\ref{solxy}) of
extremum
equations, it is easy to see that the contribution from $D$ terms to
the
determinant is negative. More precisely, taking into account
(\ref{solxy}), (\ref{Bmu}), (\ref{higgsmass}) we can see that 
$m_1^2m_2^2-m_3^4=-\fr{g_{3L}^4}{16\ov{g}^4}(m_H^2-m_{\ov{H}}^2)^2<0$ and
the
minimum condition is satisfied. For the latter, as we have already
mentioned, $m_H^2\neq m_{\ov{H}}^2$ is
crucial. This can be achieved by
renormalization between the high scale (at which soft terms are
universal) and the SUSY
scale. Therefore EWSB can naturally occur within our PGB scenario.

\subsection{Symmetries, higher order operators
and all order hierarchy}

For the scenario discussed above the $SU(2)_{cus}$
symmetry was crucial. One
may think about the origin of this symmetry. It can be obtained either
accidentally or might have a gauge origin. For example, with two discrete
symmetries ${\cal Z}_2^{(1, 2)}$ which are acting as 
${\cal Z}_2^{(1)}$: $(\ov{H}^1, H_1)\to -(\ov{H}^1, H_1)$,
${\cal Z}_2^{(2)}$: $(\ov{H}^2, H_2)\to -(\ov{H}^2, H_2)$, by imposing
the exchange symmetry ${\cal S}_2$: 
${\cal Z}_2^{(1)}\stackrel{\rightarrow}{\leftarrow} {\cal Z}_2^{(2)}$,
$\ov{H}^1 \stackrel{\rightarrow}{\leftarrow} \ov{H}^2$,
$H_1 \stackrel{\rightarrow}{\leftarrow} H_2$, the renormalizable
superpotential will have precisely the form (\ref{4Dsup}).

$SU(2)_{cus}$ can be also gauged. In fact $SU(6)\tm SU(2)$ is one of the
maximal subgroups of $E_6$ and one can assume that at high scales the
$E_6$ is broken to the $5D$ $SU(6)\tm SU(2)_{cus}$ gauge group (either by
compactification of an additional extra dimension or by some other
mechanism) and then $SU(6)$ reduces to $G_{331}$. As we see, the
origin of $SU(2)_{cus}$ symmetry can be different. Important is the form
of the superpotential (\ref{4Dsup}), which guarantees naturally light
Higgs doublets.

Let us also discuss the higher order operators and their effects. It is
easy to verify that non-renormalizable terms
of $(\ov{H}^mH_n)(\ov{H}^nH_m)$ 
would be dangerous for the hierarchy since the cancellation of
doublet's 
masses would not occur anymore. 
The reason for this is that these terms explicitly violate the
$U(6)_{Gl}$
global symmetry (the role of this symmetry was discussed above), which
arises accidentally at the renormalizable level.
To avoid quartic non-renormalizable terms, some discrete symmetries
can be applied. These can even provide an all order hierarchy. If
instead of
$SU(2)_{cus}$ one introduces a ${\cal Z}_5$ symmetry acting as
$H_{1, 2}\to H_{1, 2}$,
$\ov{H}^{1, 2}\to \om \ov{H}^{1, 2}$ ($\om =e^{{\rm i}\fr{2\pi }{5}}$), 
then the lowest allowed superpotential couplings will be
\beq
W'=M_{Pl}^3\l \fr{\ov{H}^mH_n}{M_{Pl}^2}\r^5~,
\la{highsup}
\eeq
where all possible contractions of $SU(6)$ indeces are assumed and
$M_{Pl}$($\simeq 2.4\cdot 10^{18}$~GeV) is the reduced Planck mass. In
the unbroken
SUSY limit (\ref{highsup}) gives $\lan \ov{H}^m\ran =\lan H_n\ran
=0$. However, it is easy to verify that the inclusion of soft SUSY
breaking
terms $m_{\ov{H}^{m}}^2|\ov{H}^m|^2+m_{H_m}^2|H_m|^2+A'mW'$ will give
the non-zero solutions 
$v\sim \bar v \sim M_{Pl}\l \fr{m}{M_{Pl}}\r^{1/8}$
which
for $m\sim 1$~TeV are indeed $\sim 10^{16}$~GeV. With these VEVs and
the operators
in (\ref{highsup}) the Higgs doublet pair has masses
$\sim M_{Pl}\l \fr{\bar v\cdot v}{M_{Pl}^2}\r^4\sim m$ of the wanted
size. Note that in this case, since the couplings in (\ref{highsup}) are
strongly suppressed, there is no need for a $SU(2)_{cus}$ symmetry. 
${\cal Z}_5$ guarantees the generation of non-zero $v, \bar v$ VEVs and
an all
order hierarchy. However, in this case the light Higgs doublets are not
the pseudo-Goldstones. This distinguishes this case from the model with
an $SU(2)_{cus}$ symmetry. 

As we have seen, the orbifold constructed $SU(6)$ GUT looks very promising
for
obtaining a nice hierarchy and for solving the $\mu $ problem. 
In the next section we will show that within the orbifold $SU(6)$ GUT,
extended
either with $SU(2)_{cus}$ or by ${\cal Z}_5$ symmetry, the fermion masses
are naturally generated.

\section{Quark-lepton masses}

In order to construct the fermion sector we will introduce some
$SU(6)$ states at the 5D level. We will consider here just the case of one
generation. The generalization to three families of quark-leptons will
be straightforward. 

Introduce the following states
$$
\Psi_{N=2}(15)=(15,~\ov{15})~,~~\Psi_{N=2}'(15)=(15',~\ov{15}')~,~~
$$
\beq
F_{N=2}^{(i)}(6)=(\ov{6}, ~6)^{(i)}~,~~
{F_{N=2}'}^{(i)}(6)=(\ov{6}', ~6')^{(i)}~,~~~i=1, 2~,
\la{fermions}
\eeq
where states with primes are so-called copies and are necessary if we want to
introduce matter fields in the bulk (see $1^{\rm st}$ and $3^{\rm rd}$
ref. in \cite{symbr} ). 
Two sets of sextets ($i=1,2$) are
necessary to construct an anomaly free $SU(6)$ orbifold model. In terms of
$G_{331}$, $15$ and $\ov{6}$ decompose as
$$
15=u^c(\ov{3},~1)_2+Q(3,~3)_0+E^c(1,~\ov{3})_{-2}~,~~
$$
\beq
\ov{6}=d^c(\ov{3},~1)_{-1}+{\cal L}(1,~\ov{3})_1~.
\la{dec15_6}
\eeq
The decomposition of $15'$ and $\ov{6}'$ is similar. 
Since states $\ov{15}$ and $6$ are conjugates of $15$
and $\ov{6}$ resp. their decompositions also will be conjugate 
with respect to
(\ref{dec15_6}). In terms of $G_{321}$ the states $Q$, $E^c$, ${\cal L}$
read
\beq
Q=(q,~\ov{D}^c)~,~~~E^c=(e^c,~\ov{L})~,~~~{\cal L}=(l,~\xi ) ~,
\la{decQEL}
\eeq
where $q$, $e^c$ and $l$ carry precisely the same quantum numbers as the left
handed quark doublet, right handed lepton and left handed lepton
doublet, resp.
$\ov{D}^c$ and $\ov{L}$ have quantum numbers conjugate to the right handed
down quark and left handed doublet, resp. $\xi $ is a MSSM singlet.
We ascribe the following $Z_2\tm Z_2'$ orbifold parities to the
states
\beq
\l u^c, ~Q', ~E^c, ~{\cal L}^{(1, 2)}, ~{{d^c}'}^{(1, 2)}\r \sim (+,~+)
~,~~
\l {u^c}', ~Q, ~{E^c}', ~{{\cal L}'}^{(1, 2)}, ~{d^c}^{(1, 2)}\r \sim
(-,~+)
\la{ferpar}
\eeq
and opposite parities to the mirrors. With the transformations of
gauge field fragments given in (\ref{gaugepar}) the 5D SUSY action will be
invariant.
With the parities in (\ref{ferpar}),  only the states
$u^c$, $E^c$, $Q'$, ${\cal L}^{(1, 2)}$, ${{d^c}'}^{(1, 2)}$ 
have zero modes. One can
easily verify that these states effectively constitute 
$15+2\tm \ov{6}$ representations of the $SU(6)$ gauge group, 
which then is free of anomalies.
Since at the $y=0$ fixed point we have 4D $N=1$ SUSY $G_{331}$ symmetry, the
brane Yukawa couplings also possess this symmetry.
In addition, we also have $SU(2)_{cus}$ (or ${\cal Z}_5$) symmetry. 
Making the fermion sector consistent with $SU(2)_{cus}$ we
assume that 
${\cal L}^{(1, 2)}$ and ${{d^c}'}^{(1, 2)}$ are antidoublets
of $SU(2)_{cus}$
\beq
({\cal L}^{(1)}, ~{\cal L}^{(2)})\equiv {\cal L}^{m}~,~~~
({{d^c}'}^{(1)}, {{d^c}'}^{(2)})\equiv {{d^c}'}^{m}~,~~~m=1, 2~.
\la{fermdef}
\eeq
The $G_{331}\tm SU(2)_{cus}$ invariant 4D superpotential Yukawa couplings are
\beq 
W_Y=Q'{{d^c}'}^{m}\ov{H}^n\ep_{mn}+E^c{\cal L}^{m}\ov{H}^n\ep_{mn}+
\fr{1}{M}Q'u^cH_mH_n\ep^{mn}~, 
\la{4Dyuk} 
\eeq                          
where $\ep_{mn}$ is the $SU(2)_{cus}$ invariant antisymmetric tensor.
The first two terms are responsible for the generation of down 
quark and charged 
lepton masses, respectively. The last term generates masses for up type
quarks.
The first two terms in (\ref{4Dyuk}) are also crucial for the extra 
vector-like
states to get decoupled. More precisely, substituting appropriate VEVs and
extracting appropriate states from $G_{331}$ superfields 
[see (\ref{vevH}), (\ref{MSSMpair}), (\ref{dec15_6}), (\ref{decQEL})] 
the first term in (\ref{4Dyuk}) gives
\beq
Q'{{d^c}'}^{m}\ov{H}^n\ep_{mn}\to v_0\ov{D}^c{{d^c}'}^{2}+
q'{{d^c}'}^{1}\ov{D}^2= v_0\ov{D}^c{{d^c}'}^{2}+q'{{d^c}'}^1h_d~.
\la{yukd}
\eeq
Therefore the state ${{d^c}'}^{2}$ decouples with $\ov{D}^c$ getting a 
mass
$\sim v_0$.\footnote{For VEVs of $\ov{H}^1$, $H_1$ we use $v_0$, because
the
insignificant shift between $v$ and $\bar v$ is not relevant for
decoupled vector-like states.} 
The states $q'$, ${{d^c}'}^{1}$ have Yukawa coupling 
(with the down Higgs doublet $h_d$) 
and can be identified as left handed quark doublet and
right handed down quark resp.
The second term in (\ref{4Dyuk}) gives
\beq
E^c{\cal L}^{m}\ov{H}^n\ep_{mn}\to v_0\ov{L}\hs{0.07cm}l^2+e^cl^1h_d~.
\la{yukl}
\eeq
As we see, $l^2$ decouples with $\ov{L}$ forming a massive state. States
$e^c$, $l^1$ have standard Yukawa couplings (with $h_d$) and can be
identified with the right handed lepton and left handed lepton doublet
respectively.
Finally, the last term in (\ref{4Dyuk}) leads to
\beq
\fr{1}{M}Q'u^cH_mH_n\ep^{mn}\to \fr{v_0}{M}q'u^ch_u~,
\la{yuku}
\eeq
which for 
$M\sim v_0\sim M_G$ results in $\lam_U\sim 1$ \footnote{The operator
in (\ref{yuku}) also can be obtained by exchange of appropriate
vector-like states with masses $\sim M$.}.
As we see, the usual Yukawa couplings are obtained in a natural way in
a scenario extended with $SU(2)_{cus}$ symmetry. It is easy to show that
phenomenologically required fermion sector can be also built within a
model where
$SU(2)_{cus}$ is replaced by ${\cal Z}_5$ symmetry (giving an all order
hierarchy. See sect. 2.2). For the latter case in the Yukawa
superpotential
(\ref{4Dyuk}) instead of $\ep_{mn}$ tensors some arbitrary 
couplings will
appear. It is important  that also within this scenario additional
vector
like states can be decoupled at scale $v_0$ and that at low energies we
remain
with the chiral content of MSSM. Therefore, we can conclude that within
an orbifold $SU(6)$ GUT, augmented with additional symmetries,
the fermion sector can be naturally constructed.

\section{Gauge coupling unification}

The unification of the three gauge couplings in this scenario occurs near
the
scale
$M_G\sim 10^{16}$~GeV. The first step of GUT symmetry breaking occurs at
the
compactification scale $\mu_0=1/R\stackrel{<}{_\sim }M_G$. Then below
the scale $v_0$ we have $G_{321}$ gauge symmetry with MSSM content. Since
at the scale $v_0$ some vector-like states decouple 
[see (\ref{yukd}), (\ref{yukl})], between $v_0$ and $M_G$ the b-factors
of zero modes will be
\beq
\l b_3,~b_{3L},~b_{U(1)}\r^{v_0}=\l 0,~2,~10\r ~.
\la{bv}
\eeq
Above the $\mu_0$ scale, the KK states will contribute into the
running and the corresponding factors are
$$
\l \ga_3 ,~\ga_{3L},~ \ga_{U(1)}\r =\l -6,~-2,~2\r +
6\eta \cdot \l 1,~1,~1 \r ~,
$$
\beq
\l \de_3 ,~\de_{3L},~ \de_{U(1)}\r =\l -2,~-6,~-10\r +
6\eta \cdot \l 1,~1,~1 \r ~,
\la{gade}
\eeq
where $\eta $ is number of generations living in the bulk, $\ga $ and
$\de $ correspond to KK states with masses $(2n+2)\mu_0$
and $(2n+1)\mu_0$, resp. This kind of mass spectrum is due to the
$Z_2\tm Z_2'$ orbifold parity prescriptions of (\ref{gaugepar}),
(\ref{parhiggs}), (\ref{ferpar}). Note that above the compactification
scale $\mu_0$
we consider the effective four dimensional gauge theory
with a limited tower of states \cite{ddg}.
Taking these into account, we obtain for the solution of the 1-loop RGE
\footnote{For RGEs with presence of limited KK
states see \cite{ddg}, while detailed derivation of RGEs with
intermediate scale and KK states can be found in \cite{mu3}.}
\beq
\al_3^{-1}=\fr{12}{7}\al_2^{-1}-\fr{5}{7}\al_1^{-1}+
\fr{9}{7\pi }\ln \fr{M_G}{v_0}+
\fr{6}{7\pi} (S_2-S_1)~,
\la{als}
\eeq
\beq
\ln \fr{v_0}{M_Z}=\fr{5\pi }{14}(\al_1^{-1}-\al_2^{-1})-
\fr{8}{7}\ln \fr{M_G}{v_0}+
\fr{4}{7}(S_2-S_1)~,
\la{scale}
\eeq
\beq
\al_G^{-1}=\al_2^{-1}-\fr{1}{2\pi }\ln \fr{v_0}{M_Z}-
\fr{1}{\pi }\ln \fr{M_G}{v_0}
+\fr{1-3\eta }{\pi }S_1+\fr{3-3\eta }{\pi }S_2~,
\la{alG}
\eeq
where
\beq 
S_1=\sum_{n=0}^{N} \ln \fr{M_G}{(2n+2)\mu_0}~,~~~~ 
S_2=\sum_{n=0}^{N'} \ln \fr{M_G}{(2n+1)\mu_0}~.
\la{SKK} 
\eeq  
In (\ref{SKK}), $N$ and $N'$ are the maximal numbers of appropriate KK         
states 
which lie below $M_G$, i.e. 
$(2N+2)\mu_0\stackrel{<}{_\sim }M_G$,
$(2N'+1)\mu_0\stackrel{<}{_\sim }M_G$.  
States which have masses above the $M_G$ are not relevant for unification
and therefore we truncate the sums in (\ref{SKK}) at certain $N$,
$N'$ numbers of states.

{}From (\ref{als}) we see that in order to have a reasonable value of
$\al_3$ we need $v_0\simeq M_G$ and therefore $\mu_0\sim M_G$. 
As one can easily see from (\ref{scale}),
the unification scale is close to $10^{16}$~GeV.

\section{Conclusions}

We have presented a 5D SUSY $SU(6)$ GUT construction on an 
$S^{(1)}/Z_2\tm Z_2'$ orbifold. Such a construction together with
an additional
symmetry suggests a very economical way for understanding the right
size
of $\mu $ and $B\mu $ terms. With a $SU(2)_{cus}$ symmetry the Higgs
doublets are pseudo-Goldstone Bosons emerging as a consequence of the
spontaneous breakdown of a global pseudo-symmetry. We have also shown
that alternatively the
extension with a $Z_5$ discrete symmetry leads to a natural all order
hierarchy and to the understanding of the origin of the GUT scale, which
is an
interplay of $M_{Pl}$ and the SUSY scale $m\sim 1$~TeV. In both
types of extensions $\mu $ and $B\mu $ emerge dynamically after SUSY
breaking.

{\bf Acknowledgement} 

F.P.C. is supported by Funda\c c\~ ao de Ci\^ encia e Tecnologia (grant
SFRH/BD/4973/2001).

\bibliographystyle{unsrt}

\begin{thebibliography}{99}  


\bibitem{kawa}
Y. Kawamura, Prog. Theor. Phys. 105 (2001) 999; 
{\it ibid.} 105 (2001) 691.
%


\bibitem{dtpdecay}
G. Altarelli, F. Feruglio, hep-ph/0102301;
A. Kobakhidze, hep-ph/0102323.


\bibitem{symbr}
L. Hall, Y. Nomura, Phys. Rev. D 64 (2001) 055003;
M. Kakizaki, M. Yamaguchi, hep-ph/0104103;
A. Hebecker, J. March-Russel, Nucl. Phys. B 613 (2001) 3;
R. Barbieri, L. Hall, Y. Nomura, hep-th/0107004; 
C. Csaki, G. Kribs, J. Terning, hep-ph/0107266;
N. Maru, hep-ph/0108002;
Q. Shafi, Z. Tavartkiladze, hep-ph/0108247;
T. Li, hep-ph/0108120; hep-th/0110065.


\bibitem{orbSO10}
T. Asaka, W. Buchm\"uller, L. Covi, hep-ph/0108021;
L. Hall, Y. Nomura, T.  Okui, D. Smith, hep-ph/0108071;  
R. Dermisek, A. Mafi, hep-ph/0108139;
N. Haba, T. Kondo, Y. Shimizu, hep-ph/0112132, hep-ph/0202191.


\bibitem{mu1}
G. Giudice, A. Masiero, Phys. Lett. B 206 (1988) 480.
%
\bibitem{mu2}
S. Dimopoulos, G. Dvali, R. Rattazzi, Phys. Lett. B 413 (1997) 336;
E.J. Chun, Phys. Rev. D 59 (1999) 015011; P. Langacker, N. Polonsky,
J. Wang, Phys. Rev. D 60 (1999) 115005;
T. Han, D. Marfatia, R.J. Zhang, Phys. Rev. D 61 (2000) 013007;
K. Choi, H.D. Kim, Phys. Rev. D 61 (2000) 015010;
Q. Shafi, Z. Tavartkiladze, Nucl. Phys. B 580 (2000) 83.
%
\bibitem{mu3}   
F. Paccetti Correia, M.G. Schmidt, Z. Tavartkiladze, hep-ph/0204080.
%
\bibitem{pgbsu61}
K. Inoue, A. Kakuto, T. Takano, Progr. Theor. Phys. 75 (1986) 664;
A. Anselm, A. Johansen, Phys. Lett. B200 (1988) 331;
Z. Berezhiani, G. Dvali,
Sov. Lebedev Institute Reports 5 (1989) 55.
%
\bibitem{pgbsu62}
R. Barbieri, G. Dvali, A. Strumia, Nucl. Phys. B391 (1993) 487;
R. Barbieri, G. Dvali, M. Moretti, Phys. Lett. B312 (1993) 137;
R. Barbieri et al.,
Nucl.Phys. B432 (1994) 49;
Z. Berezhiani, Phys. Lett. B355 (1995) 481;
Z. Berezhiani, C. Csaki, L. Randall, Nucl. Phys. B444 (1995) 61;
G. Dvali, S. Pokorski, Phys. Rev. Lett. 78 (1997) 807;
Q. Shafi, Z. Tavartkiladze, 
Nucl. Phys. B 552 (1999) 67;
Nucl. Phys. B 573 (2000) 40;
B. Bajc, I. Gogoladze, R. Guevara, G. Senjanovic,
Phys. Lett. B 525 (2002) 189.
%
%
\bibitem{pgbsu3}
%
%
G. Dvali, Q. Shafi, Phys. Lett. B 326 (1994) 258; B 339 (1994) 241;
G. Lazarides, C. Panagiotakopoulos, Q. Shafi,
Phys. Lett. B 315 (1993) 325.
%
\bibitem{radbr_pgb}
$2^{nd}$ ref. in \cite{pgbsu62};
C. Csaki, L. Randall, Nucl. Phys. B 466 (1996) 41;
B.Ananthanarayan, Q. Shafi, Phys. Rev. D 54 (1996) 3488.
%
\bibitem{cus} 
%
G. Dvali, Phys. Lett. B 324 (1994) 59.
%
\bibitem{cusext}
%
I. Gogoladze, A. Kobakhidze, Z. Tavartkiladze,
Phys. Lett. B 372 (1996) 246;
A. Kobakhidze, Phys. Lett. B 391 (1997) 335;  
Z. Tavartkiladze, Phys. Lett. B 392 (1997) 360;
I. Gogoladze, hep-ph/9612365.
%
\bibitem{sugra}
%
R. Barbieri, S. Ferrara, C. Savoy, Phys. Lett. B 119 (1983) 123;
H.P. Nilles, M. Srednicki, D. Wyler, Phys. Lett. B 120 (1983) 346;
L.J. Hall, J. Lykken, S. Weinberg, Phys. Rev. D 27 (1983) 2359.
%
\bibitem{radbr}
%
L. Ibanez, G. Ross, Phys. Lett. B 110 (1982) 215;
K. Inoue et al., Prog. Theor. Phys. 68 (1982) 927;
Y. Okada, M. Yamaguchi, T. Yanagida, Phys. Lett. B 62 (1991) 54;
H. Haber, R. Hempfing, Phys. Rev. Lett. 66 (1991) 1815;
J. Ellis, G. Ridolfi, F. Zwirner, Phys. Lett. B 257 (1991) 83;
R. Barbieri, M. Frigeni, F. Caravaglios, Phys. Lett. B 258 (1991) 167. 
%
%

\bibitem{ddg}
%
K. Dines, E. Dudas, T. Gherghetta, Nucl. Phys. B 537 (1999) 47. 

\end{thebibliography}

\end{document}